\documentclass[aps,nofootinbib,floatfix,showpacs,preprintnumbers,twocolumn]{revtex4} 
\usepackage{graphicx}
\usepackage{bm}
\usepackage{mathrsfs}
\usepackage{float}
\usepackage{tabularx}
\usepackage{amsmath,amsfonts,amssymb}

\input epsf

\graphicspath{{/Users/lstrigar/Dropbox/NeutrinoBackground/sterile/figures/}}

\def\lsim{\mathrel{\raise.3ex\hbox{$<$\kern-.75em\lower1ex\hbox{$\sim$}}}}
\def\gsim{\mathrel{\raise.3ex\hbox{$>$\kern-.75em\lower1ex\hbox{$\sim$}}}}

\def\cmm2{{\,\rm cm^{-2}}}
\def\cm2{{\,{\rm cm}^2}}
\def\cmm3{{\,{\rm cm}^{-3}}}
\def\gcmm3{{\,{\rm g\,cm^{-3}}}}

\def\fun#1#2{\lower3.6pt\vbox{\baselineskip0pt\lineskip.9pt
  \ialign{$\mathsurround=0pt#1\hfil##\hfil$\crcr#2\crcr\sim\crcr}}}

\def\be{\begin{equation}}
\def\ee{\end{equation}}
\def\bea{\begin{eqnarray}}
\def\eea{\end{eqnarray}}

\begin{document}

\title{Comparing readout strategies to directly detect dark matter}
\author{J.~Billard}\email{j.billard@ipnl.in2p3.fr} \affiliation{IPNL, Universit\'e de Lyon, Universit\'e Lyon 1, CNRS/IN2P3, 4 rue E. Fermi 69622 Villeurbanne cedex, France}

\begin{abstract}

Over the past decades, several ideas and technologies have been developed to directly detect weakly interacting massive particle (WIMP) from the galactic halo. All these detection strategies share the common  goal of discriminating a WIMP signal from the residual backgrounds. By directly detecting WIMPs, one can measure some or all of the observables associated to each nuclear recoil candidates, such as their energy and direction. In this study, we compare and examine the discovery potentials of each readout strategies from counting only (bubble chambers) to directional detectors (Time Projection Chambers) with 1d-, 2d-, and 3d-sensitivity. Using a profile likelihood analysis, we show that, in the case of a large and irreducible background contamination characterized by an energy distribution similar to the expected WIMP signal, directional information can improve the sensitivity of the experiment by several orders of magnitude. We also found that 1d directional detection is only less effective than a full 3d directional sensitivity by about a factor of 3, or 10 if we assume no sense recognition, still improving by a factor of 2 or more if only the energy of the events is being measured.

\end{abstract}
\maketitle

\section{Introduction}
\label{sec:intro}
An ever increasing body of evidence supports the existence of cold dark matter (CDM) as a major contribution to the matter budget of the Universe. On the largest scale, cosmological measurements~\cite{komatsu} tightly constrain the CDM relic density whereas on a local scale, measurements of the rotation curves of spiral galaxies, including the Milky Way, indicates that they should be embedded in a dark matter halo~\cite{persic,klypin}. A leading candidate for this dark matter is a yet-to-be-discovered weakly interacting massive particle (WIMP) which could directly interact with detectors based on Earth leading to keV-scale nuclear recoils. Direct detection experiments are now probing well-motivated extensions to the standard model which naturally predict dark matter candidates~\cite{Jungman,Bertone,Strigari}. However, as dark matter detectors are rapidly improving in sensitivity~\cite{Demarteau:2014pka}, they will encounter the neutrino background, at which point Solar, atmospheric, and diffuse supernova neutrinos will interfere with a potential dark matter signal~\cite{neutrinoJocelyn, neutrinoLouie, neutrinoCRESST, neutrinoBillard, neutrinoRuppin}. Moreover, the recent controversy in the low-mass WIMP region $\sim \mathcal{O}$(10 GeV/c$^2$), where several dark matter hints are inconsistent with null results~\cite{Cushman:2013zza}, highlights the need for additional discrimination power between WIMP events and backgrounds in order to clearly authenticate a genuine WIMP signal.\\

Several ideas and detector readouts have been developed over the past decades that allows to detect keV-scale nuclear recoils as produced by $\mathcal{O}$(10-1000)~GeV/c$^2$ WIMPs. As of today, the main categories of experiments are: cryogenic semiconductor detectors~\cite{Agnese:2014aze, Armengaud:2013vci, Aalseth:2014jpa, Angloher:2014myn}, single- (liquid) and dual-phase (liquid/gas) Argon or Xenon Time Projection Chambers (TPC)~\cite{Akerib:2013tjd, Aprile:2013doa, Amaudruz:2014nsa, Agnes:2014bvk}, bubble chambers operating such that they are only sensitive to nuclear recoils~\cite{Behnke:2013sma, Archambault:2012pm, SIMPLE}, and low-pressure gaseous TPC aiming at measuring both the energy and the track of the recoiling nucleus~\cite{Battat:2014van, Ahlen:2010ub, newage, Santos:2013hpa}. Each of these detection techniques share the common goal of discriminating a potential WIMP signal from residual backgrounds by either comparing the different  amount of energy released in scintillation, ionization and heat, and/or using pulse shape discrimination. Depending on the readout strategy being considered, direct detection experiments can have access to the number of WIMP candidates contained in the data set, their recoil energies and directions. In this study we want to compare the different readout strategies by evaluating their discovery potential in various experimental conditions, especially when the data set is contaminated with some irreducible backgrounds.

This paper is organized as follows. In Sec.~\ref{sec:rates}, we briefly review the dark matter rate calculations with a particular emphasis on the directionality of WIMP induced events in both the galactic and detector-based coordinates. We then discuss the analysis methodology used to compare the discovery reach associated to each readout strategies in Sec.~\ref{sec:analysis} and discuss the detector configurations used in our simulations in Sec.~\ref{sec:detector}. Finally, we present our results in Sec.~\ref{sec:results} and conclude in the last section.

\section{Direct detection of dark matter}
\label{sec:rates}

Direct dark matter detection aims at detecting  elastic scattering between a WIMP from the galactic halo and the detector material. The differential event rate as a function of both the recoil energy ($E_r)$ and direction in the lab frame ($\Omega_r$) is given by
\begin{equation}
\frac{\mathrm{d}^2R}{\mathrm{d}E_r\mathrm{d}\Omega_r} = MT\times\frac{\rho_0\sigma_0}{4\pi m_{\chi}\mu^2_N}F^2(E_r)\hat{f}(v_{\text{min}},\hat{q};t),
\label{directionalrate}
\end{equation}
with $m_{\chi}$ the WIMP mass, $\mu_N$ the WIMP-nucleus reduced mass, $\rho_0$   the local dark matter density, and $\sigma_0$ is the WIMP-nucleus cross section. $F(E_r)$ is the nuclear form factor that corresponds to the Fourier transform of the nuclear density distribution and describes the loss of coherence for nonzero momentum transfer. Note that its expression depends on the nature of interaction: spin independent (SI) or spin dependent (SD)~\cite{lewin}. $M$ and $T$ are, respectively, the detector mass and the exposition time of the experiment. $v_{\text{min}}$ is the   minimal WIMP velocity required to produce a
nuclear recoil of energy $E_r$ and $\hat{q}$ is its direction in the detector (lab) frame at the sidereal time $t$.
Finally, $\hat{f}(v_{\text{min}},\hat{q})$ is the three-dimensional Radon transform of the WIMP 
velocity distribution $f(\vec{v})$ which has the following expression for the standard halo model (SHM)~\cite{gondolo}:
\begin{equation}
\hat{f}(v_{\text{min}},\hat{q};t) = \frac{1}{(2\pi\sigma_v^{2})^{1/2}}\exp{\left[-\frac{\left[v_{\text{min}} + \hat{q}.\vec{v}_{\rm lab}\right]^2}{2\sigma^2_v}\right]}.
\end{equation} 
where $\sigma_v = v_0/\sqrt{2}$ is the isotropic velocity dispersion, $v_0$ is the circular velocity taken to be equal to 220~km/s in the following and  $\vec{v}_{\rm lab}$ is the lab velocity vector with respect to the Galactic rest frame. The latter will be further detailed below.
As one can see from this equation, the directional rate is maximum when $\hat{q}.\vec{v}_{\rm lab} = -v_{\rm min}$ (maximum along a ring, see~\cite{ring}), if $v_{\rm min}<v_{\rm lab}$, or when $\hat{q} = -\vec{v}_{\rm lab}$ otherwise (dipole feature, see~\cite{spergel}). Therefore, we can see that the recoil direction is strongly correlated with the lab's motion in the galactic frame, which will lead to a nonambiguous authentication of a genuine WIMP signal~\cite{Annedisco, Billarddisco}.

Note that we have considered the SHM model even though recent results from N-body simulations indicate that this Maxwell-Boltzmann assumption might be an oversimplification~\cite{Vogelsberger:2008qb,Kuhlen:2009vh,Mao:2012hf}. Further, the existence of many possible substructures such as streams and a dark disk could create distinct features in the velocity distribution~\cite{Nezri, Ling, Bruch, Read,O'Hare:2014oxa}, and then have important implications in the interpretation of dark matter data. However, since the goal of this paper is to compare different readout strategies within a consistent picture to what is being used to derive existing constraints,  we will consider this simplified halo model.

\begin{figure*}
\begin{center}
\includegraphics[width=0.34\textwidth]{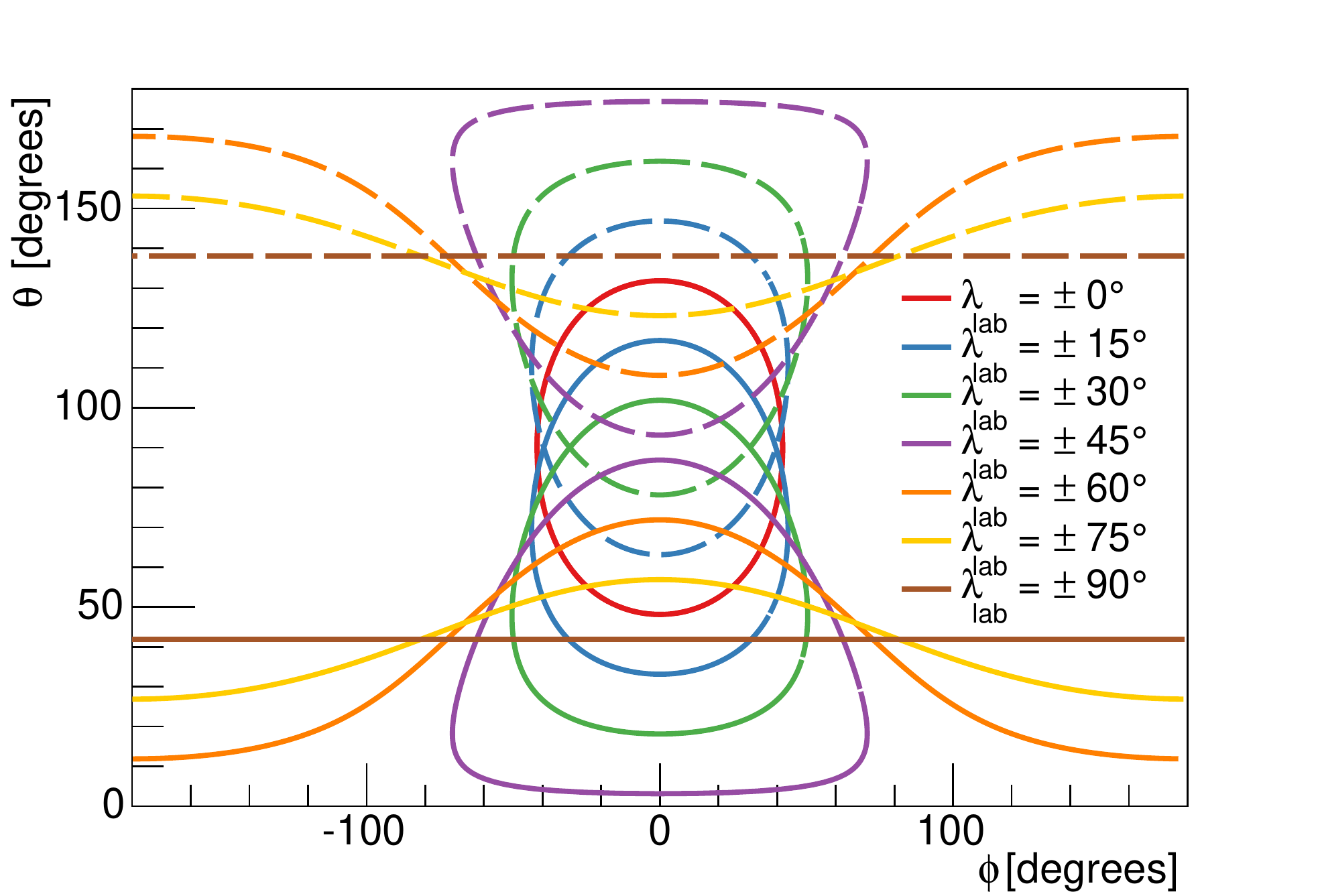}
\hspace{-0.6cm}
\includegraphics[width=0.34\textwidth]{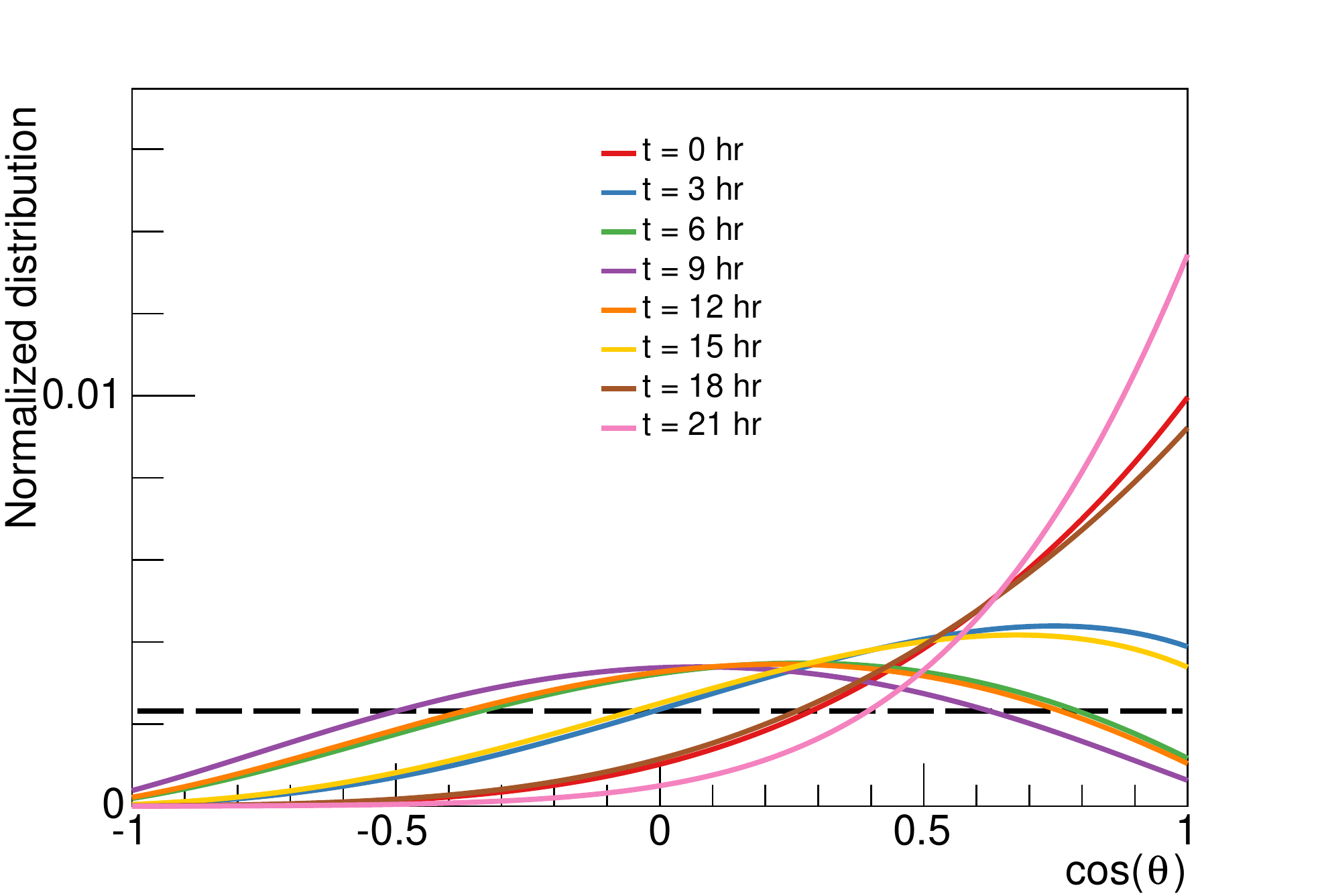}
\hspace{-0.6cm}
\includegraphics[width=0.34\textwidth]{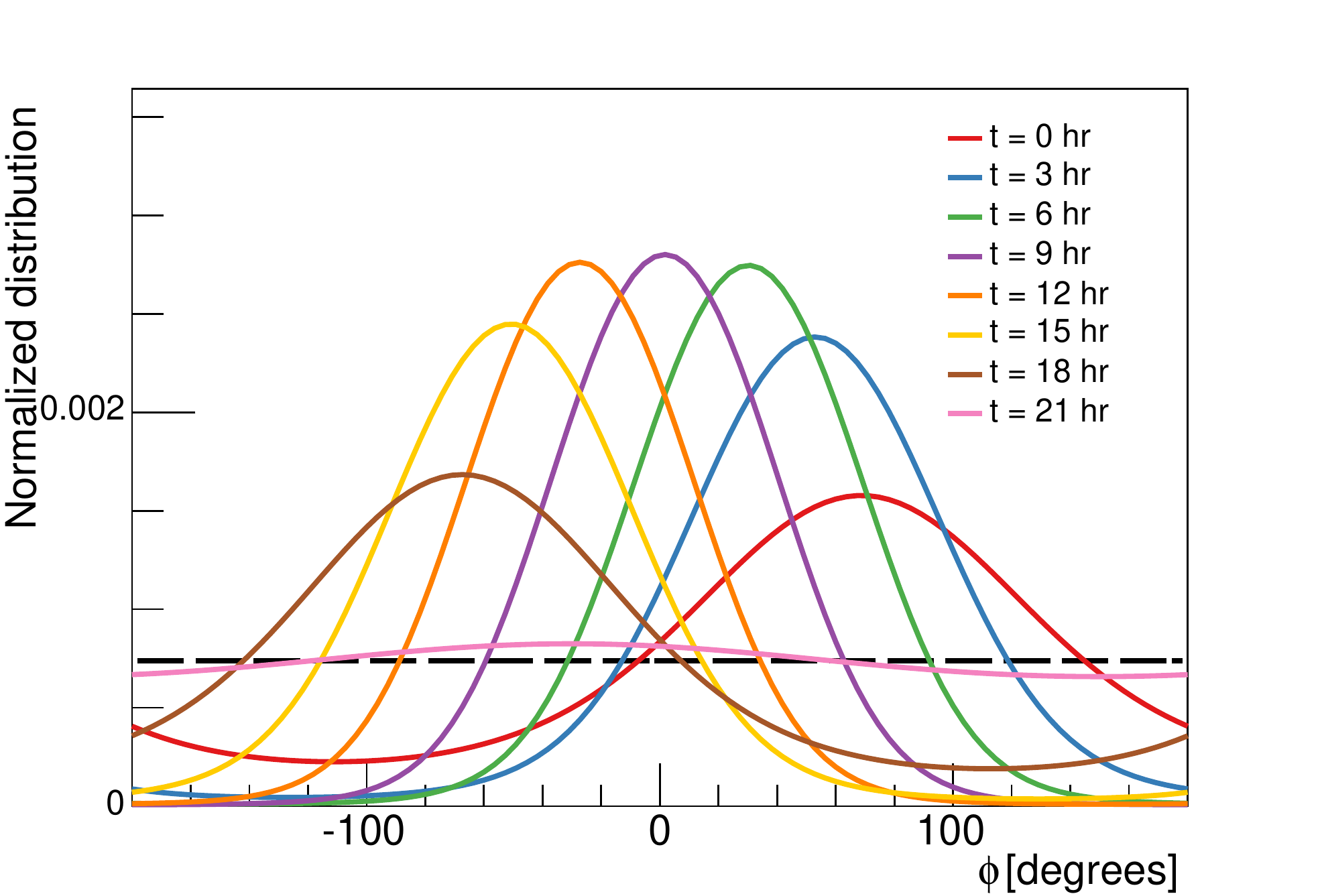}
\end{center}
\caption{Left: Evolution of the direction pointing toward  the constellation Cygnus over a sidereal day in the detector's coordinates. Middle and right:  Angular distributions of WIMP induced nuclear recoils at 50 keV in the detector frame along the $\theta$ (middle) and $\phi$ (right) angles over a sidereal day. We assumed a Xe-based experiment located in Modane and a WIMP mass of 50 GeV/c$^2$. The black dashed line correspond to an isotropic distribution.}
\label{fig:sidereal}
\end{figure*}

As discussed above, the directional event rate is strongly correlated to the lab velocity vector in the galactic rest frame. It is given by the sum of the rotation of the solar system around the galactic center $\vec{v}_{\rm GalRot}$, the Solar System peculiar velocity $\vec{v}_{\rm Solar}$, the Earth's revolution $\vec{v}_{\rm EarthRev}$ and the Earth's rotation $\vec{v}_{\rm EarthRot}$. Interestingly, $\vec{v}_{\rm EarthRev}$ and $\vec{v}_{\rm EarthRot}$ are, respectively, responsible for the annual~\cite{annual} and diurnal~\cite{daily} modulation effects. However, their contributions are negligible compared to $\vec{v}_{\rm GalRot}$. In the following we will therefore only consider this dominant contribution, {\it i.e.}, the detector velocity is such that $\vec{v}_{\rm lab} = \vec{v}_{\rm GalRot}$ where ${v}_{\rm GalRot}$ = 220 km/s along the  galactic $\hat{y}_G$ axis pointing toward the constellation Cygnus at ($l$ = 90$^\circ$, b = 0$^\circ$). As we will be considering both 1d and 2d directional readouts in the following, it is necessary to perform coordinate transformations from the galactic to the detector rest frame in order to estimate the directional signals as observed by such detectors~\cite{daily}. In the following, we will consider a detector with the $\hat{x}$,  $\hat{y}$, and $\hat{z}$ axes pointing toward the North, the West and the Zenith directions respectively. For a detector located at a latitude $\lambda_{\rm lab} = 90^\circ$, {\it i.e.}, at the North Pole, $\hat{z}$ is aligned with the spin axis of the Earth. Finally, in the detector frame, the direction of a recoil is given by the angle $\theta$ and $\phi$ defined as such:
\begin{equation}
\hat{q} = \sin\theta\cos\phi \hat{x} + \sin\theta\sin\phi\hat{y} + \cos\theta \hat{z}
\label{eq:angles}
\end{equation}

Figure~\ref{fig:sidereal} (left panel) shows the evolution of the direction of $\hat{v}_{\rm lab}$ in the  detector's coordinates system as a function of sidereal time for $\lambda_{\rm lab}$ varying from -90$^\circ$ to 90$^\circ$. The middle and right panels illustrate the angular distributions of WIMP induced nuclear recoils at 50 keV as observed in the detector's frame, at different sidereal time, along the $\theta$ and $\phi$ angles. These calculations have been performed considering a Xe-based experiment located in the underground laboratory of Modane (LSM) ($\lambda_{\rm lab} = 45.2^\circ$) with a WIMP mass of 50 GeV/$c^2$. As one can see from Fig.~\ref{fig:sidereal}, due to the Earth's revolution, the directional dark matter event rate is highly dependent on the sidereal time and on the latitude at which the experiment is located. It is also worth noticing that an experiment only sensitive to the $\theta$ angle (1d directionality) or the $\phi$ angle (2d directionality) should still be able to distinguish between a dark matter signal and an isotropic background.

\section{Analysis Methodology}
\label{sec:analysis}

In this section we introduce the analysis methodology used in the following in order to assess the discovery potential of each readout strategies. Discovery limits were first introduced in~\cite{billard1}, in the context of directional detection, and are defined such that if the true WIMP model lies above such limit, then an experiment has a 90\% chance to get at least a $3\sigma$ discovery. The calculation of the discovery significance is based on the standard profile likelihood ratio test statistic~\cite{cowan1} where the likelihood function is defined as,
\begin{align}
\mathscr{L}(\sigma_{\chi-n},R_b) &= \frac{e^{-(\mu_\chi+\mu_b)}}{\rm N!}\nonumber \\
& \times \prod_{i=1}^{\rm N}\left[\mu_\chi f_\chi(\vec{q}_i;t_i) + \mu_b f_b(\vec{q}_i)\right]\times\mathscr{L}(R_b),
\end{align}
where $\mu_\chi$, $\mu_b$ and $N$ are, respectively, the expected number of WIMP and background events, and the total number of observed events. Note that $\mu_b = R_b \times M T$ where $R_b$ is the background rate. $f_\chi$ and $f_b$ are the unit normalized event distributions for the WIMP and the background contributions. $\vec{q}_i$ corresponds to the set of observables $\{E_r, \theta_r, \phi_r \}$, depending on the readout considered, associated to each observed nuclear recoil, and $t_i$ is its sidereal time. Finally, $\mathscr{L}(R_b)$ is the likelihood function related to the background rate of the experiment. The latter is parametrized as a Gaussian distribution with a standard deviation given by the relative uncertainty on the expected background rate $\sigma_{R_b}$.

The profile likelihood ratio corresponds to a hypothesis test against the null hypothesis $H_0$ (background only) and the alternative $H_1$, which includes both background and signal, while  incorporating any type of systematic uncertainties such as the background normalization. In the context of a discovery significance estimate,  we are interested in testing the background only hypothesis ($H_0$) on the data and try to reject it using the following likelihood ratio:
\begin{equation}
\lambda(0) = \frac{\mathscr{L}(\sigma_{\chi-n} = 0,\hat{\hat{R}}_b)}{\mathscr{L}(\hat{\sigma}_{\chi-n},\hat{R}_b)},
\end{equation}
where $\hat{\hat{R}}_b$ denotes the values of $R_b$ that maximizes $\mathscr{L}$ for the specified $\sigma_{\chi-n} = 0$, {\it  i.e.}, we are profiling over  $R_b$ which is considered as a nuisance parameter. As discussed in Ref.~\cite{cowan1}, the test statistic $q_0$ is then defined as:
\begin{equation}
q_0 = \left\{
\begin{array}{rrll}
\rm & -2\ln\lambda(0)	&	\ \hat{\sigma}_{\chi-n} > 0 \\
\rm & 0  		& 	\ \hat{\sigma}_{\chi-n} < 0. 
\end{array}\right.
\end{equation}
As one can deduce from such test, a large value of $q_0$ implies a large discrepancy between the two hypotheses, which is in favor of a discovery interpretation.  Following Wilk's theorem, $q_0$ asymptotically follows a half $\chi^2$ distribution with one degree of freedom (see Ref.\cite{cowan1} for a more detailed discussion). In such a case, the significance $Z$ in units of sigmas of the detection is simply given by $Z = \sqrt{q^{\rm obs}_0}$.

It is worth noticing that such analysis methodology is particularly relevant to recent dark matter searches analyses and was first introduced by the XENON10 collaboration~\cite{Aprile:2011hx}. Today, many experiments, such as LUX~\cite{Akerib:2013tjd}, CDMS~\cite{CDMSJulien,Agnese:2014xye}, and CoGeNT~\cite{Aalseth:2014jpa}, are  also using likelihood approaches as their background estimates and related systematics, based on simulations and calibration data, are getting more and more reliable. The great interest of using likelihood analyses is that they offer a possible interpretation of the data, assuming that both the signal and background models are accurate, and lead to the best dark matter sensitivity.

\section{Detector configuration}
\label{sec:detector}

In this section, we describe the characteristics of the detector considered hereafter. Unless otherwise stated, we will consider a Xe-based experiment located in Modane ($\lambda_{\rm lab} = 45.2^\circ$) with a nuclear recoil energy range from 5 to 100 keV. As mentioned above,  we will assume that the detector's reference frame is such that $\hat{x}$, $\hat{y}$, and $\hat{z}$ are, respectively, pointing toward the North, West, and Zenith directions and that the $\theta$ and $\phi$ angles are defined as in Eq.~(\ref{eq:angles}). In all results shown below, we will assume a WIMP mass of 50 GeV/c$^2$ which leads to a mean recoil energy of $\sim$10 keV and is usually where most Xenon-based experiments have their best WIMP sensitivities. Neglecting the impact of the form factor which varies from target to target, equivalent directional signals can be found by adjusting the energy range for each target~\cite{Billarddisco}. For example, a similar WIMP-induced recoil distribution would be obtained for $^{19}$F target with $3 \leq E_r \leq 60$ keV. 
In the following, we will consider 5 types of detector readout strategies:
\begin{itemize}
\item Counting experiment: the detector is only able to measure a total number of events above some threshold. This is the case for the bubble chamber experiments~\cite{Cushman:2013zza},  which adjust their operating pressure to nucleate a single bubble from a nuclear recoil.
\item Energy: this category corresponds to the bulk part of  direct detection experiments where only the energy of the events is being measured~\cite{Cushman:2013zza}. The kinetic energy of the recoiling nucleus is, in most cases, derived from multiple component measurements: heat/ionization for semiconductor cryogenic detectors, ionization and scintillation such as Xe- and Ar-based dual-phase TPCs, and heat/scintillation for cryogenic scintillating crystals. 
\item Energy + 3d: The energy and the track of the recoiling nucleus are fully measured. This is the ultimate readout strategy that fully exploit the information from the expected WIMP signal. Current directional experiments are using low-pressure gaseous TPC in order to get few mm tracks associated to $\mathcal{O}(10)$ keV nuclear recoils. For 3D sensitive directional detectors, the track is measured by sampling over time the 2 dimensional projection of the ionization-induced electron cloud on a pixelized anode (see~\cite{Ahlen} and references therein). The drawback of such detection techniques is that current directional detectors only have about $\sim$0.1 kg of target material and are therefore not yet competitive with the above mentioned experiments.
\item Energy + 2d: The energy is being measured as well as the 2d projection of the recoil track onto the (x,y) plane of the detector to have access to the $\phi$ angle. This is the case of most current directional detectors that do not have access to a time sampling of the track projection on the anode~\cite{Ahlen}.
\item Energy + 1d: The energy is being measured as well as the 1d projection of the track along the z-axis to have access to the $\theta$ angle. This would be the case for a dual-phase Xe or Ar TPC looking for columnar recombination where the electric field would be aligned along the z-axis, taken to be pointing toward the zenith direction. Note that such detection technique has only very recently been suggested by D. Nygren~\cite{nygren} but has generated an increasing interest. So far, we have no experimental evidence that such effect would be measurable at the keV-scale as it requires both tracks sufficiently long and high energy resolutions on both the ionization (S2) and scintillation (S1) channels. So far, only the SCENE collaboration has seen a possible deviation, at  $\sim$60 keV for Argon recoils, in the S1 signal between tracks selected to be perpendicular or parallel to the electric field~\cite{Cao:2014gns}. Although several ideas on how to enhance this columnar recombination effect exist~\cite{nygren}, we are not yet at the stage  where such detectors are  able to realize a highly efficient directional detection of dark matter. However, it is certainly worth looking into the discovery potential of an hypothetical 1d directional sensitivity especially as such detectors would not suffer from low-mass target material, unlike gaseous directional detectors. Also, it is worth noticing that this work is the first addressing the discovery potential of such 1d directional readout.
\end{itemize}

For all readouts, unless otherwise stated, we will consider the case of an ideal detector as this study is meant to compare the ultimate reach of each of these detection techniques. Moreover, a comprehensive study of the impact of realistic detector limitations, such as energy and angular resolutions and energy thresholds,  has been done in~\cite{billard1} where a similar profile likelihood technique were used.

\begin{figure*}
\begin{center}
\includegraphics[width=1.05\columnwidth, height=7cm]{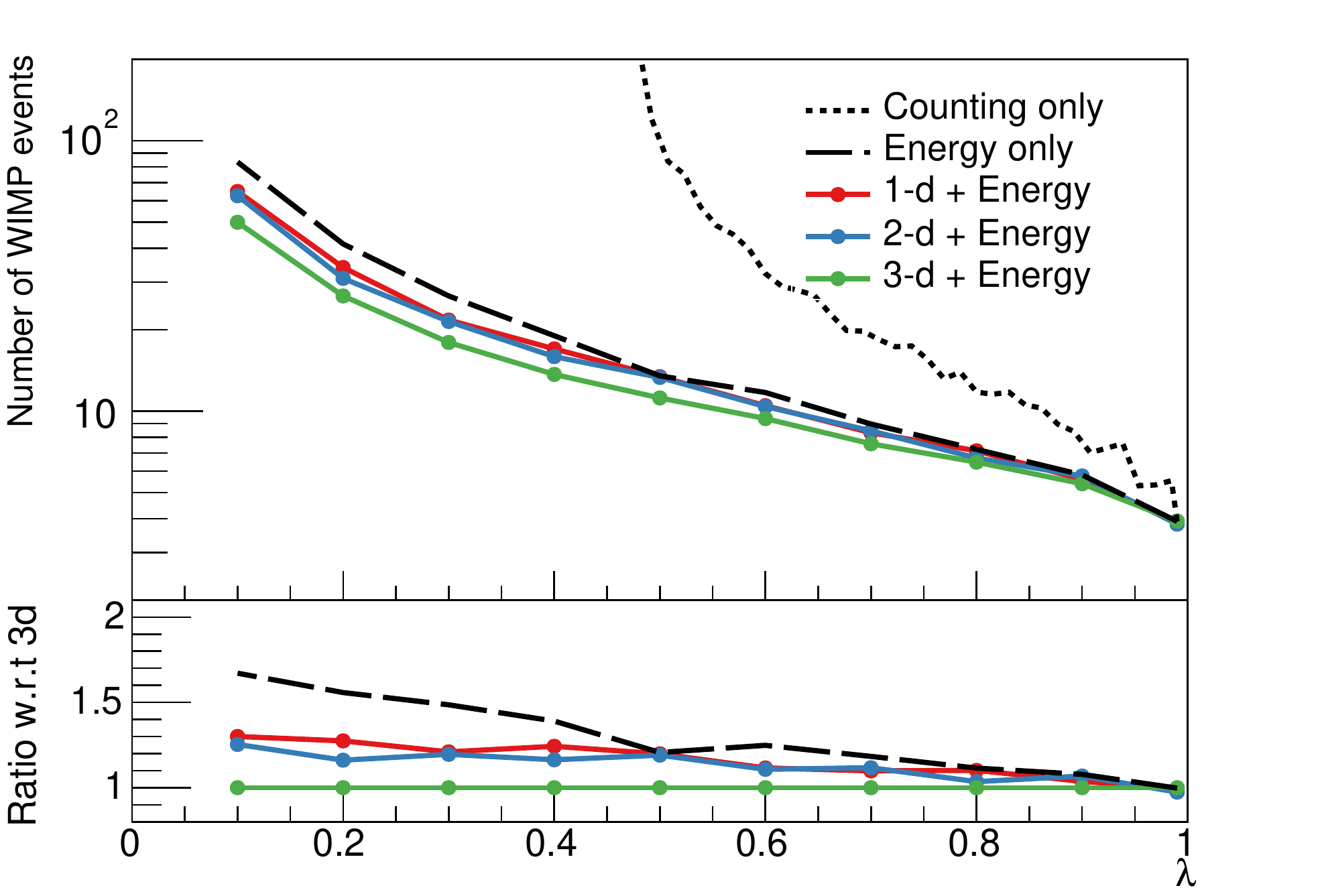}
\hspace{-1cm}
\includegraphics[width=1.05\columnwidth, height=7cm]{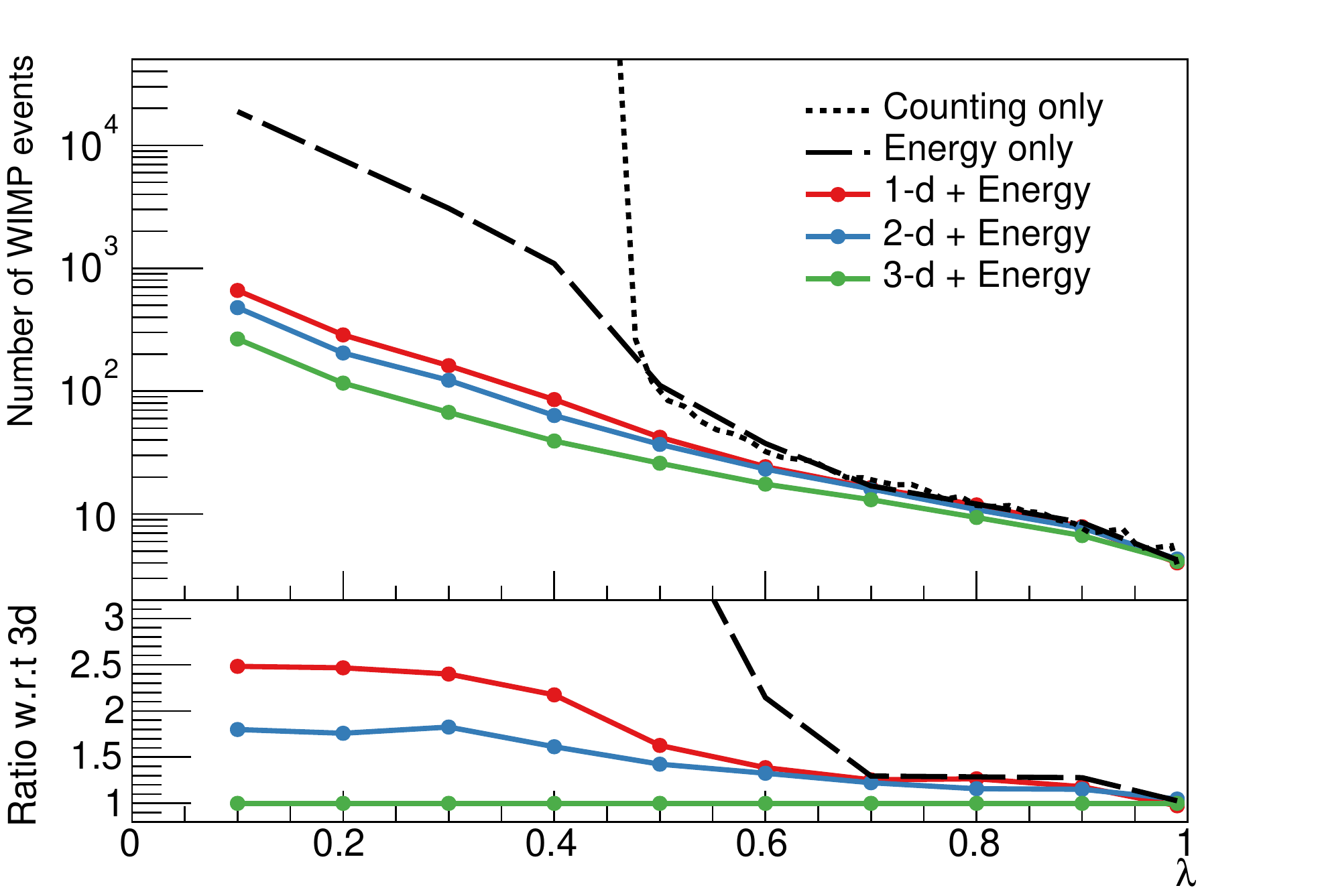}
\end{center}
\caption{Number of WIMP events required to get a 3$\sigma$ detection in 90\% of the experiment as a function of the signal purity $\lambda$ for all different types of readout: counting only (black dotted line), Energy only (black dashed line), Energy + 1d, 2d and 3d directionality (red, blue, and green solid lines, respectively). We considered a Xe-based experiment located in Modane with a recoil energy range from 5 keV to 100 keV and a 50 GeV/c$^2$ WIMP mass. Left (right) panel assumes an  isotropic background with an exponential energy distribution with a slope of 100 (10) keV. Small fluctuations in the results are due to the finite size of the Monte Carlo simulations of 1000 samples.}
\label{fig:lambda}
\end{figure*}

\section{Results}
\label{sec:results}

This section is dedicated to presenting our results in the comparison of the different type of readouts that are sensitive to the different observables of a potential WIMP signal: number of WIMP induced nuclear recoils, their energy and their direction. Throughout this study, we considered an isotropic background with an exponential energy distribution with a slope of either 10~keV or 100~keV. The motivation for choosing an isotropic residual background comes from the fact that residual neutrons in an underground laboratory have been estimated to closely follow an isotropic distribution~\cite{mei}. The relative uncertainty on the background rate estimate is taken to be equal to $\sigma_{R_b} = 20$\% which is of similar order to what was considered in the recent LUX analysis~\cite{Akerib:2013tjd} and is comparable to uncertainties on the ultimate neutrino background~\cite{neutrinoBillard, neutrinoRuppin}.\\

Figure~\ref{fig:lambda} shows the evolution of the number of WIMP events required to get a 3$\sigma$ detection significance in 90\% of the experiments as a function of the signal purity $\lambda = \mu_\chi/(\mu_\chi + \mu_b)$ for all 5 readout strategies. The left and right panels correspond to a background energy slope of 100 keV and 10 keV respectively. Note that in the 10 keV case (right panel), the energy distribution of the background and the one induced by a 50 GeV/$c^2$ WIMP are very similar. From Fig.~\ref{fig:lambda} (both panels) one can see that the number of required WIMP events to achieve a significant dark matter discovery increases as the signal purity decreases. Also, one can notice that the ``counting'' readout, which cannot distinguish a genuine WIMP event from the background, saturates around $\lambda = 0.4$. This comes from the fact that below such value, the background contribution is so important that because of its associated normalization uncertainty $\sigma_{R_b}$ it is not possible to get a significant discovery anymore. Note that the breaking point $\lambda = 0.4$  only depends on the relative uncertainty on the background normalization considered. Indeed, if $\sigma_{R_b}$ increases, this breaking point tends to values closer to $\lambda = 1$. For example, with $\sigma_{R_b}$ = 50\% we found this breaking point to be equal to $\lambda = 0.7$.

As one can see from Fig.~\ref{fig:lambda} (left panel), where we expect a reasonable discrimination in energy between the WIMP signal and the background, that the ``Energy only'' and the four directional readouts give comparable sensitivities. As a matter of fact, at the lowest signal purity $\lambda = 0.1$, we find that the number of required events to reach a high significance discovery using only the information in energy is about a factor 1.6 higher than if this information is combined with a full 3d track measurement. This means that the 50 GeV/c$^2$ WIMP energy distribution is sufficiently different from the one of an exponential background with a slope of 100 keV that directional information does not completely overcome the ``Energy only'' readout used by most direct detection experiment. Additionally one can notice that 2d and 1d readouts are roughly comparable and about only a factor 1.3 worse, at the lowest signal purities, than a 3d readout.

From Fig.~\ref{fig:lambda} (right panel), one can see that the results are significantly different if we consider a background model characterized by an energy distribution that mimics very well the expected WIMP induced energy distribution. Indeed, in such a case, measuring only the energy of the events does not bring discrimination power between the two hypotheses. As a matter of fact, one can see that the ``Energy only'' scenario is equivalent to the ``Counting only'' scenario down to $\lambda = 0.5$. However, below such signal purity, one can see that the number of required WIMP events jumps to about 1000 at $\lambda = 0.4$ but does not saturate unlike the ``Counting only'' case. This is explained by the fact that the two spectra, WIMP and background, are not exactly alike, especially in the tail as the WIMP energy distribution is not exactly an exponential distribution. Therefore with sufficiently high statistics, one can discriminate between the WIMP and the background hypotheses. This is why the ``Energy only'' case can still perform a high significance detection of dark matter below $\lambda = 0.4$ but at the price of a huge increase in exposure. Unlike the energy spectrum, no background can potentially mimic exactly the directional distribution of WIMP induced nuclear recoils. Therefore, no matter the level of signal purity, there is always sufficient discrimination between the WIMP and background events to reach a highly significant dark matter detection even with only a 2d or a 1d readout, as shown in Fig.~\ref{fig:lambda}. The interest of such study is that it gives us a sense on how each directional readout strategies perform compare to each other when the discrimination along the energy alone is negligible. As one can see, all three directional readout strategies are about 2 orders of magnitude below the ``Energy only'' case. Furthermore, one can see that at the lowest signal purity, the 3d readout outperforms the 1d and 2d readouts by about a factor 1.8 and 2.5, respectively. \\

\begin{figure}[t]
\includegraphics[width=\columnwidth]{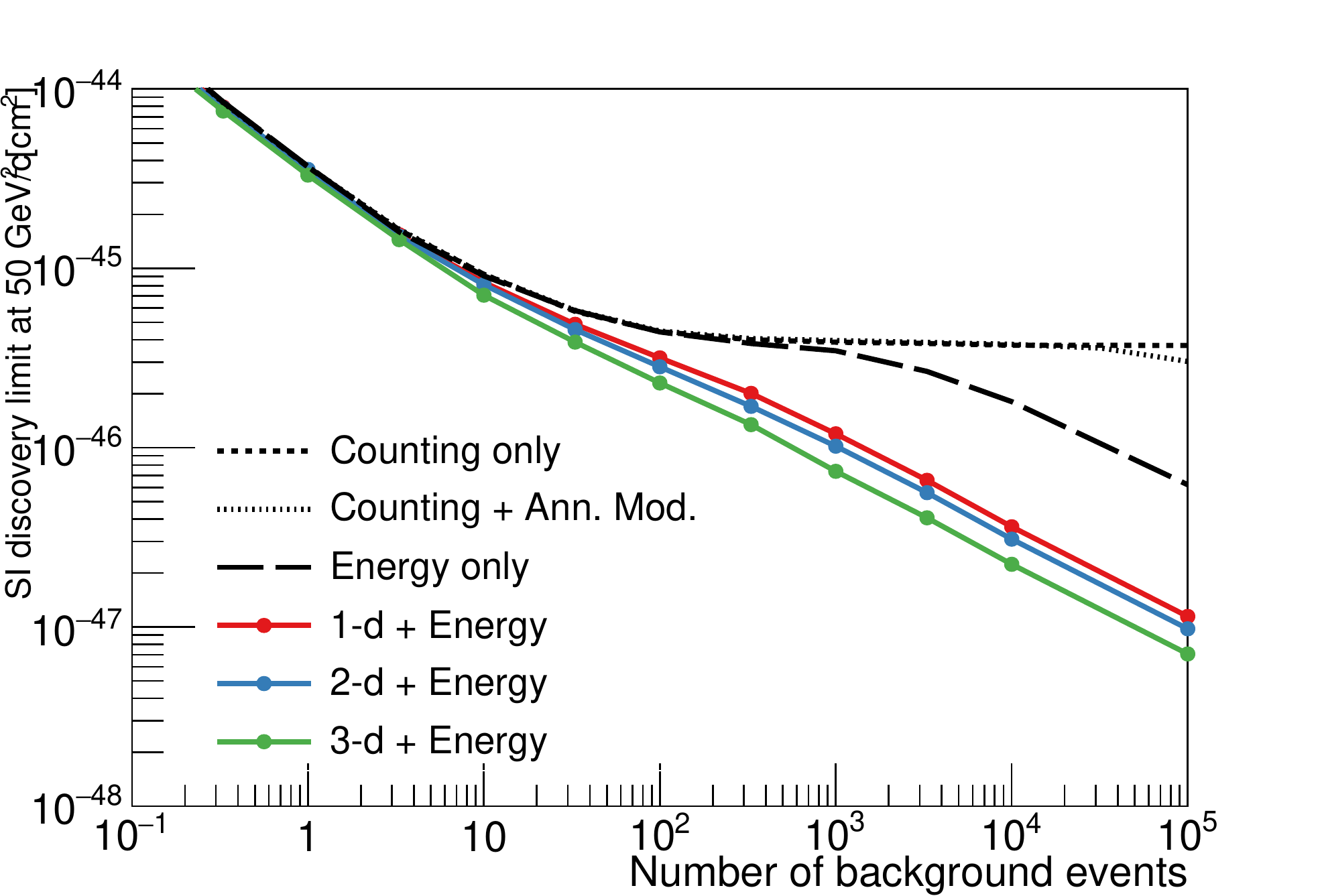}
\caption{Evolution of the discovery limit for a SI interaction for a 50 GeV/c$^2$ WIMP mass as a function of the number of expected background events (exposure) for all different types of readout: counting only (black dotted line), counting only including the effect of annual modulation (black short-dotted line), Energy only (black dashed line), Energy + 1d, 2d and 3d directionality (red, blue, and green solid lines respectively). We considered a Xe-based experiment located in Modane with a recoil energy range from 5 keV to 100 keV, a background rate of 100 events/ton/year, a systematic uncertainty of $\sigma_{R_b}$ = 20\%, and an exponential slope of 10 keV. Small fluctuations in the results are due to the finite size of the Monte Carlo simulations of 1000 samples.}
\label{fig:expo}
\end{figure}

Figure~\ref{fig:expo} shows the evolution of the discovery limit for a SI interaction and for a fixed WIMP mass of 50~GeV/$c^2$ as a function of the exposure for all types of readouts. One can see that for low exposures, where the number of expect background events is very low (below 0.1), the discovery sensitivity evolves as $1/MT$ similarly for all readout strategies. Once the experiments start to be sensitive to the background contamination, we can observe different behaviors. The ``Counting only'' case saturates for an expected number of background events higher than 100 as no background subtraction is possible. For completeness, the effect of annual modulation on the ``Counting only'' case has also been investigated and is shown to only improve on the sensitivity at very high exposures due to the small ($\sim$ 2\%) modulation in the event rate expected for such WIMP mass, recoil energy range and background rate uncertainties. Note that no improvements were found by taking into account annual modulation effects with the other readout strategies.  In the ``Energy only'' case, the sensitivity has a similar behavior than the ``Counting'' case as the background and WIMP energy spectra are very similar. However, for expected number of background events higher than 1000, the small differences in the two energy distributions allow the ``Energy only'' readout to perform some background subtraction hence leading to a sensitivity scaling as $1/\sqrt{MT}$. Note that this behavior is very similar to what was observed in the case of the neutrino background~\cite{neutrinoRuppin}. Interestingly, one can see that all three directional readouts have sufficient discrimination as they directly get to a background subtraction mode, characterized by a sensitivity scaling as  $1/\sqrt{MT}$, when the background becomes important enough. Finally, from Fig.~\ref{fig:expo} we can deduce that to probe a $10^{-46}$ cm$^2$ SI cross section, one would need 100 times more exposure with an ``Energy only'' readout than with a fully directional detector. A 1d and 2d readouts would, respectively, only require 2.5 and 1.8 times more exposure than the 3d case. Also, we can see that such cross section is out of reach for a counting experiment only, under such background considerations.

This clearly highlights the interest of pursuing the construction of large scale directional detectors, especially when the background can mimic the energy distribution of a  putative WIMP signal. Note that this is expected to happen when upcoming experiments will be sensitive to the neutrino background which can almost perfectly mimic a WIMP signal, see~\cite{neutrinoBillard, neutrinoRuppin}. In such case, only the directional information will help at probing dark matter models beyond this neutrino background~\cite{Grothaus:2014hja}. These results also suggest that, even though it is not as optimal as a full 3d readout, a large scale 1d or 2d sensitive detector could clearly extract a WIMP signal from a highly background contaminated data sample and therefore lead to significant dark matter discoveries. \\

\begin{figure}[t]
\includegraphics[width=\columnwidth]{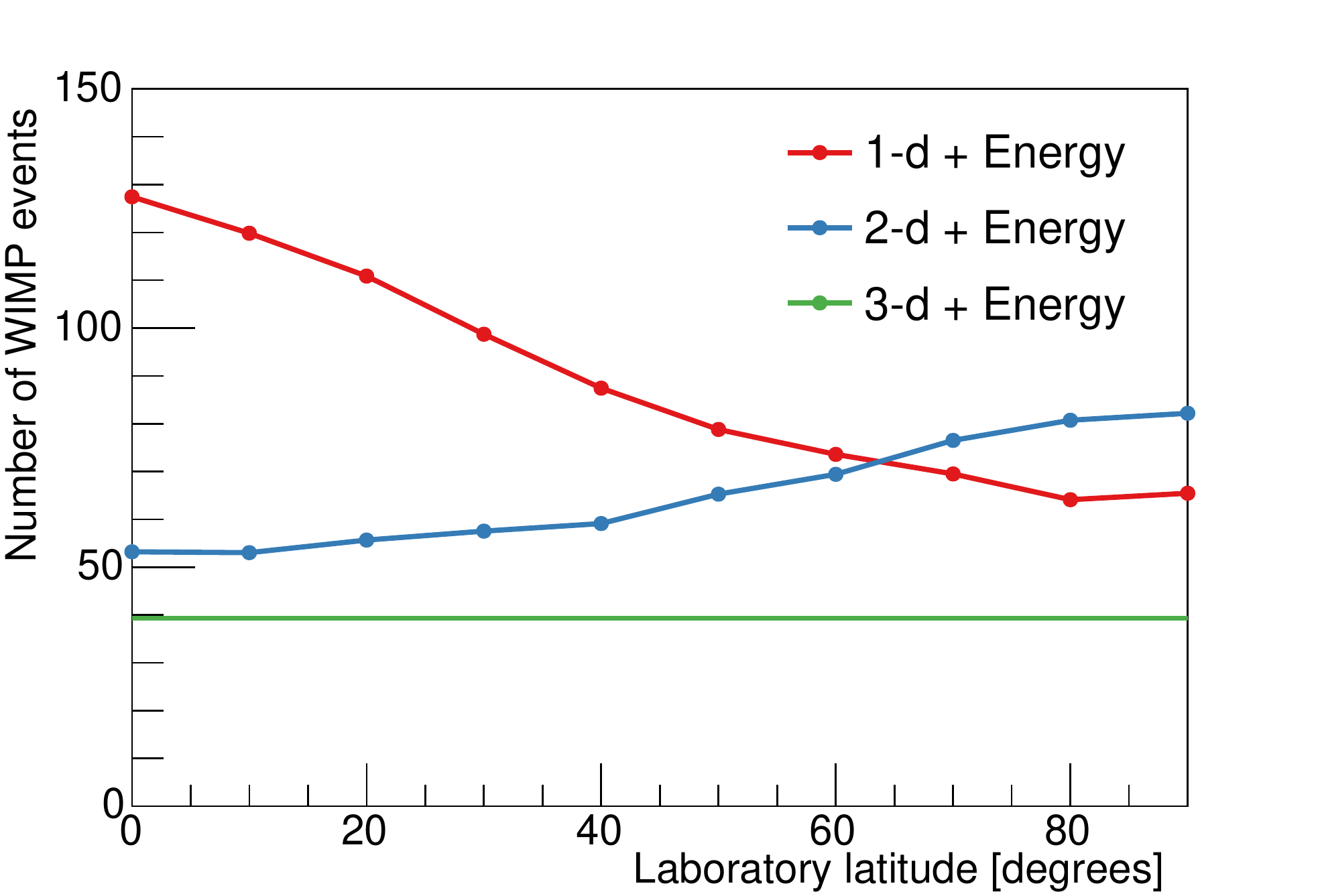}
\caption{Number of WIMP events required to get a 3$\sigma$ detection in 90\% of the experiment as a function of the detector latitude $\lambda_{\rm lab}$ for all different directional readouts: Energy + 1d, 2d and 3d directionality (red, blue, and green solid lines, respectively). We considered a Xe-based experiment with a recoil energy range from 5 keV to 100 keV, a 50 GeV/c$^2$ WIMP mass, a signal purity of $\lambda = 0.4$, and an exponential energy distribution for the isotropic background of 10 keV. Small fluctuations in the results are due to the finite size of the Monte Carlo simulations of 1000 samples.}
\label{fig:latitude}
\end{figure}

Figure~\ref{fig:latitude} shows the evolution of the number of required WIMP events  to reach a high significance detection of dark matter as a function of the detector's  latitude $\lambda_{\rm lab}$ for the three directional readouts and for similar conditions as in Fig.~\ref{fig:lambda} (right panel) at $\lambda = 0.4$. Note that the results are similar for the North and the South hemispheres (see Fig.~\ref{fig:sidereal}), hence $\lambda_{\rm lab}$ is only taken between 0$^\circ$ and $90^\circ$. As one can see, the required exposure varies with the detector latitude by about a factor of 2 and 1.4 for a 1d and a 2d readout, respectively. Also, the optimal latitude for a 1d (2d) readout, when considering the detector orientation as discussed above, is at $\lambda_{\rm lab} = 90^\circ$ ($\lambda_{\rm lab} = 0^\circ$) when the z-axis is parallel (perpendicular) to the Earth's spin axis. For both cases, such optimal orientations allow for an improved time evolution of the angular distributions. Note that the truly optimal detector's orientation would be to have the z-axis maintained parallel and perpendicular to $\vec{v}_{\rm lab}$ for the 1d and 2d readouts, respectively.  However, such configurations would require the detector to be mounted onto a gyroscope, which seems challenging. Interestingly, most underground laboratories have latitudes $|\lambda_{\rm lab}|$ between 55$^\circ$ and 30$^\circ$, highlighting the need for optimizing the detector's orientation to improve its sensitivity. It is worth noticing that these conclusions are in  agreement with previous works studying 2d directional readouts~\cite{green1,Copi:2005ya}.  \\

\begin{figure}[t]
\includegraphics[width=\columnwidth]{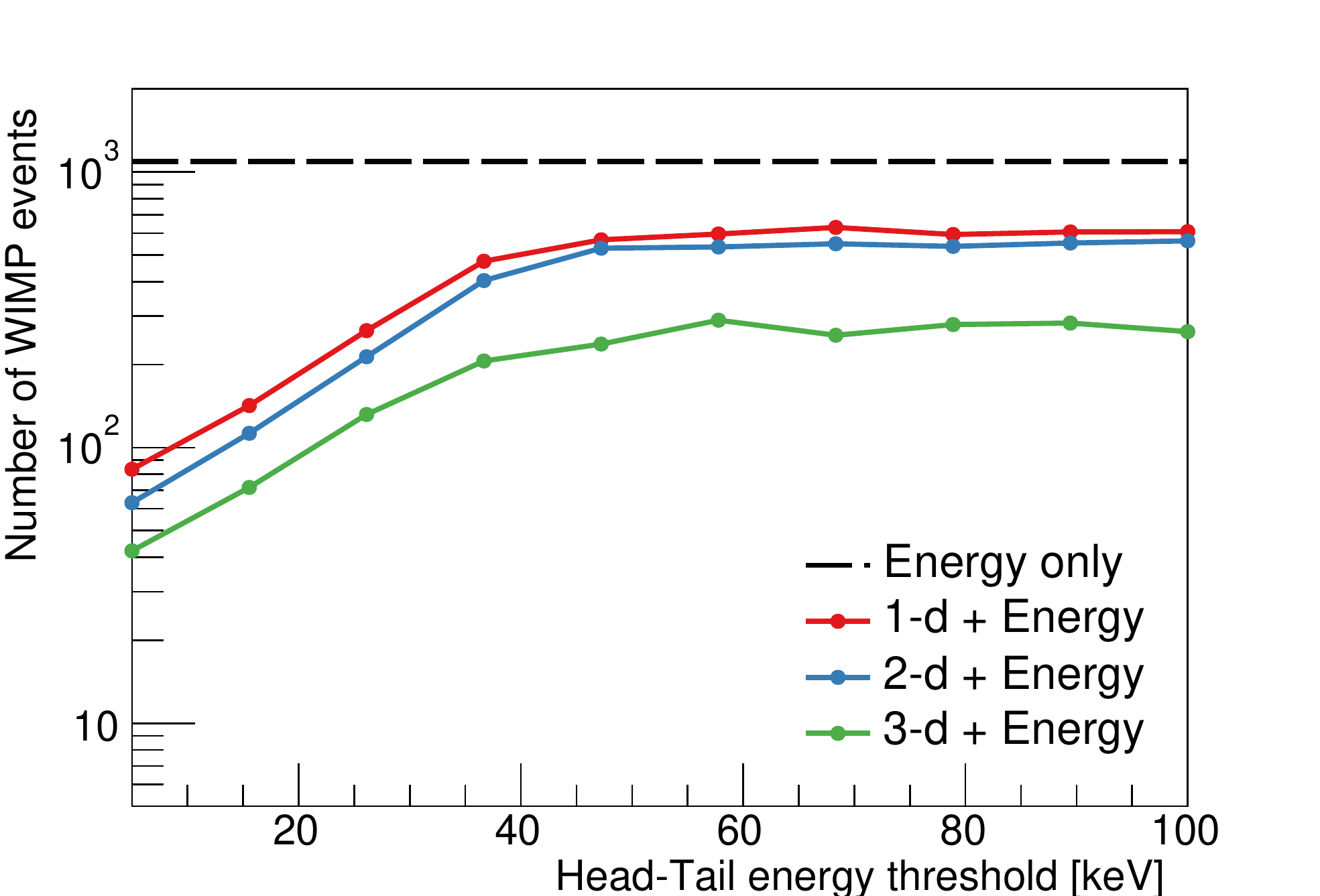}
\caption{Number of WIMP events required to get a 3$\sigma$ detection in 90\% of the experiment as a function of the Head-Tail energy threshold for the three directional readouts: 1d (red), 2d (blue) and 3d (green). Also shown for comparison is the case of reading out only the energy of the events (black dashed line). We considered a Xe-based experiment located in Modane with a recoil energy range from 5 keV to 100 keV, a 50 GeV/c$^2$ WIMP mass, a signal purity of $\lambda = 0.4$, and an exponential energy distribution for the isotropic background of 10 keV. Small fluctuations in the results are due to the finite size of the Monte Carlo simulations of 1000 samples.}
\label{fig:HT}
\end{figure}

As a final study, we explored the impact of being able to recover the sense of the recoil ($\pm\hat{q}$), the so-called ``Head-Tail'' effect, on the discovery potential of directional experiments. We will only focus on this particular experimental consideration as it has been shown to have the largest impact on directional sensitivity in previous studies~\cite{green1,Copi:2005ya}. Figure~\ref{fig:HT} shows the evolution of the required number of WIMP events to have a 90\% probability to reach a 3$\sigma$ discovery significance as a function of the ``Head-Tail'' threshold for similar conditions as in Fig.~\ref{fig:lambda} (right panel) at $\lambda = 0.4$. The ``Head-Tail'' efficiency curve is defined such that the sense recognition is 100\% (0\%) efficient above (below) its threshold. Therefore the full and null ``Head-Tail'' capabilities correspond to the case where the ``Head-Tail'' threshold is equal to 5 keV and 100 keV, respectively. For completeness, the results obtained by the different directional readouts are compared to the ``Energy only'' case, shown as the black dashed line. As one can see for all three directional readouts, the number of required WIMP events increases with the ``Head-Tail'' threshold and eventually flattens out around 50 keV as almost no WIMP events lie above such recoil energy. The loss in sensitivity when no ``Head-Tail'' capabilities is considered is about a factor of 6 for a 3d readout and about a factor of 8 for both the 2d and 1d readouts. These results suggest that not having sense recognition down to the lowest recoiling energies is not excessively penalizing, especially for heavy WIMP masses, as all three directional readouts outperform the ``Energy only'' case by at least a factor of 2. However, for light WIMPs, it is obviously very important to be able to lower down as much as possible the ``Head-Tail'' threshold.\\

The results presented above are quantitatively different than from previous studies comparing 3d and 2d readouts~\cite{green1,Copi:2005ya}. Indeed, in these works it was shown that a loss of one order of magnitude in sensitivity was observed when going from a 3d to a 2d readout. These previous works were based on nonparametric statistical tests considering the marginalized angular distribution over time and energy unlike here where we suggest to use all available observables, including time and energy to improve the discrimination between the WIMP and background hypotheses (see Fig.~\ref{fig:sidereal}). We have also shown that, even when the energy on its own is not a good discriminator (see Fig.~\ref{fig:lambda} right panel), a 2d or 1d directional readout can still bring sufficient discrimination power to reach a high significance discovery without needing a huge increase in exposure compared to the ideal 3d case. This is explained by the fact that even if the energy on its own is not a great discriminator, it is still of great help to discriminate WIMP versus background by looking at the correlations between the energy and the direction of the track which are very different between these two hypotheses.

\section{Conclusion}

We have examined the discovery reach of all possible direct detection readout strategies that  have access to some or all of the observables that one can extract from the expected WIMP signal: the number of WIMP-induced recoils, their energy and directions. We have shown that directionality is particularly valuable if the residual background energy spectrum can closely mimic the one from a possible WIMP signal. Note that this is expected to happen when experiments will be sensitive enough so that solar, atmospheric, and diffuse supernova neutrinos will interfere with a potential WIMP signal~\cite{neutrinoBillard, neutrinoRuppin}. In such case, we have shown that even a 1d directional detector could greatly improve on the discovery potential of an "Energy only" experiment. Also, we have shown that 1d and 2d  readouts are only less efficient than a full 3d readout by about a factor of 3 and $\sim$10 with and without sense recognition capabilities. However, as 1d readouts based on columnar recombination are not expected to be sensitive to the sense of recoiling nucleus they will only improve on the ``Energy only'' readout by about a factor of 2. But on the other hand, such detection technique could easily have sufficient target material to be competitive with current large-scale dark matter detectors and suffer much less from the irreducible neutrino background. It is however worth emphasizing that, even though a 1d directional sensitivity could greatly help at convincingly detecting dark matter, a 3d readout would be necessary to characterize its velocity distribution in our vicinity and begin the era of ``WIMP astronomy''~\cite{billardMCMC,lee,Lee:2014cpa}.\\

{\bf Acknowledgments}-- The author would like to thank Anne Green and Fr\'ed\'eric Mayet for their comments and suggestions on this work and also Alexandre Juillard and Jules Gascon for fruitful discussions. The author is grateful to the LABEX Lyon Institute of Origins (ANR-10-LABX-0066) of the Universit\'e de Lyon for its financial support within the program ``Investissements d'Avenir'' (ANR-11-IDEX-0007) of the French government operated by the National Research Agency (ANR).

\end{document}